\documentclass[aps,prper,twocolumn,superscriptaddress,letterpaper,longbibliography,nofootinbib]{revtex4-1}

\usepackage{graphicx}
\usepackage{mdframed}
\usepackage{framed}
\usepackage{indentfirst}
\usepackage{natbib}
\usepackage{amsmath,amssymb}
\usepackage{subfigure}
\usepackage{enumitem}
\usepackage{multirow}
\usepackage{epstopdf}
\usepackage{comment}
\usepackage{censor}
\usepackage[T1]{fontenc}	
\usepackage[latin9]{inputenc}	
\usepackage{geometry}		
\usepackage[above,below]{placeins}	
\usepackage{times}

\usepackage{hyperref}  
\hypersetup{colorlinks=true,urlcolor=blue,citecolor=blue,linkcolor=blue}   
\usepackage{enumitem}          
\setlist{nosep}                 

\begin{document}

\begin{titlepage}


\title{Perspectives from Physics Graduate Students on Their Experiences in NSF Research Experiences for Undergraduates}


\author{Jonan-Rohi S. Plueger}
\affiliation{Department of Physics, University of Colorado, 390 UCB, Boulder, CO 80309}

\author{Bethany R. Wilcox}
\affiliation{Department of Physics, University of Colorado, 390 UCB, Boulder, CO 80309}

\begin{abstract}
National Science Foundation (NSF) funded Research Experiences for Undergraduates (REUs) are explicitly intended to reach minoritized students in STEM and those who have few research opportunities. Many undergraduates are encouraged to seek them out, but their actual efficacy is not well-established, and the out-of-state travel required for many attendees may prove a significant barrier for the very students REUs wish to reach. We interviewed physics graduate students who attended REUs as undergrauates, focusing on how the REUs benefitted them, barriers they faced attending REUs, and their relationship with their REU mentors. Interviewees reported benefits that aligned with the NSF goals: skills, enculturation, and knowledge they had not received in their undergraduate institutions. They also reported financial barriers they faced which they were able to overcome due to their financial privilege. Participants also reported widely varying experiences with their mentors. Some mentors did and some did not meet their mentees where they were at in their career and skill levels. Some students did not know how to approach their mentors with their questions or needs.
  \clearpage
\end{abstract}

\maketitle

\end{titlepage}

\section{\label{sec:intro}Introduction \& Background}

The National Science Foundation (NSF) funds a collection of summer undergraduate research experiences (UREs) called Research Experiences for Undergraduates (REUs). REUs generally recruit STEM students from outside their host institution and often from out-of-state, pay them a stipend, and employ them in research for roughly 10 weeks over the summer. The NSF aims for REUs to reach underrepresented demographics in STEM and to reach students who have limited research opportunities available to them, such as students in two-year colleges \cite{nsfREUsolicitation}. The first author, a commuter student at a small school, personally benefited from NSF REUs, which were integral to his development as a scientist and his application to graduate school.

Nevertheless, students in the very demographics the NSF intends to reach may experience serious challenges attending REUs otherwise intended for them. Going to an REU usually involves packing your belongings, stepping out of the context of your normal life, and travelling to another state for ten weeks. Even with the travel and housing funds many REUs provide, moving to an REU is a significant life decision. The preliminary work of this paper begins a larger project to study the accessibility of physics REUs to underrepresented populations in physics and to optimize REUs for their benefit.

UREs take many forms: directed research, independent study, extracurricular employment in a lab, course-based undergraduate research experiences (CUREs), summer research programs, and more. While there is a general body of work on UREs, only a few aspects of physics-specific UREs have been studied so far. Quan and Elby \cite{quan2016, quan2018} have studied students' self-efficacy and experience of the physics community of practice in their research programs. Lewandowski and collaborators are currently studying several implementations of CUREs in physics \cite{oliver2023, werth202201, werth202202}. REUs, as programs that represent summer research, are an important and largely missing part of the discussion around physics UREs.

Our research is partly motivated by Alaee, et al. \cite{alaee2023, alaee2022} who broke ground on studying physics REUs. In their 2022 paper on remote REUs, they said ``REUs often require students to travel long distances'' and suggested that ``[this] may pose significant challenges for students with geographic constraints, family obligations, or health concerns that limit travel.'' It is not known how many prospective students face these barriers.

Besides the work above, there is an overall lack of physics-specific UREs in the literature--for example, there are none in the cited works of a 2015 review paper \cite{linn2015,mentorshipSTEMM}, which looks at papers studying the benefits of UREs. REUs are often recommended to students as highly valuable experiences--at the first author's small undergraduate institution, professors strongly advised him to find REUs to get research experience they could not provide. But there is not yet research-based evidence to back up the perceived high value of physics REUs.

Based on the research studying mentorship in STEM, we suspect that REU mentors are crucial to developing their students' physics identity, attitudes toward physics research, professional career, and skills \cite{thiry2011}. REUs occupy a short time-frame and target students with limited experience, which may impact REU mentors' priorities and strategies.

To study these research gaps, we gathered qualitative data from interviews concerning the following research questions: (1) What benefits might students receive from physics REUs that are relevant to their future careers? (2) What barriers do students face in attending REUs? and (3) What is the relationship between REU students and their mentors?
\vspace{-0.2\baselineskip}
\section{\label{sec:context} Methods}

We recruited physics graduate students for focus group interviews in which we asked open-ended questions about aspects of their experience with REUs. We chose the focus group format to generate a wide variety of responses and to draw out situations where one interviewee's response might provoke another's. In practice, this happened a number of times; interviewees offered a diverse range of responses, but also frequently remembered details of their experiences after hearing the responses of others in the group. In case students might have felt uncomfortable sharing some information with others in the room, we offered the option to discuss such information in private after the interview or in a one-on-one setting. No student opted to take this option. We are confident that each student communicated their unique experiences without undue influence from the responses of others. 

The focus groups were conducted in four sessions, each of which was attended by 3 of the 12 interviewees. We asked them four questions: (1) what REUs the students attended and when, (2) what barriers or challenges they faced when deciding whether to attend their REUs, (3) what benefits they derived from their REUs, and (4) how they would describe their relationship with their mentor(s) during their REUs.

The graduate students were volunteers recruited from a large physics graduate program at an R1 university, based on whether they had attended an REU as undergraduates or if they had been accepted to one or more REUs but were not able to attend any of them for some reason. While we solicited participation from this latter population, we ultimately received volunteers only from students who were able to attend REUs. Eight of our 12 interviewees responded to a short demographic survey, indicating that they were: 3 white cis women, 1 white trans woman, 1 white nonbinary person, and 3 Latino men.
Two students indicated that they came from small undergraduate institutions. At least five students had an REU early in their undergraduate career (directly after their freshman or sophomore year), but none indicated that they attended a two-year college. Students' stage in graduate school varied widely.

We opted to focus this preliminary study on graduate students because they are uniquely qualified to comment on several aspects of our research questions. Most of our interviewees indicated that they are currently engaged in graduate-level research. Ref. \cite{linn2015}, a review paper on the benefits of UREs, criticised self-report studies on the basis that students may wrongly over- or under-value their gains. Graduate students, however, can reflect more critically on the actual role REUs played in their development as a scientist, a perspective which Harsh et al. \cite{harsh2011} utilized in a 2011 longitudinal study.

The self-selection of graduate students nevertheless imposes limitations on our sample. The graduate students included in this study were admitted into a large and competitive graduate program at an institution with a largely affluent and predominantly white student population. These students are, thus, less likely to come from low-income backgrounds and less likely to be severely impacted by any socioeconomic barriers which we discover through this and future studies. Additionally, if someone's REU experience was detrimental to them, they may be less likely to appear in a sample of graduate students at a highly competitive program. For these reasons, the barriers we present should in no way be considered comprehensive. However, in cases where these students report a barrier they faced, we can reasonably assume that this is a barrier which could affect a wide variety of less affluent students as well. Moreover, aspects of REUs that were considered effective for students in our sample may not generalize as effective aspects for all REU students. Nevertheless, we can at least lay out a set of real possibilities for barriers REU students can face and benefits they can derive.

Interviews were recorded and then transcribed using Otter.ai. For this exploratory study, we chose to emergently code the transcripts (using Dedoose) so that results would arise from student responses as opposed to the researchers' prior expectations. We looked for and codeed common themes in students' experiences of their mentors, benefits they reported from their REUs, and factors that they said affected or could have affected their decision to attend a given REU, as well as any level of importance they might have ascribed to each factor. In what follows, we will describe themes that arose from coding and support those themes with relevant quotes by interviewees.

\vspace{-0.4\baselineskip}
\section{\label{sec:results}Results \& Discussion}

Students discussed their answers to questions concerning the benefits they gained from their REUs, barriers or challenges they faced in attending their REUs, and their relationship with their mentor(s). As discussed in more detail in the sections below, the benefits that students described aligned with the NSF's intention to reach students with limited access to research. Issues surrounding pay and out-of-pocket costs were a frequent challenge for students, and while these issues did not prevent these students from attending their REUs, they illuminate the influence of privilege on whether someone else might be able to. Students' experiences with mentorship varied; many provided training, a few were outright disdainful of the student, and in many cases, there was a mismatch of expectations between students and mentors which prevented communication. We will refer to students by pseudonyms representing their gender and ethnicity.

\vspace{-0.4\baselineskip}
\subsection{\label{sec:di}Benefits from attending REUs}

We told students, ``Now that you're a graduate student doing research, we'd like you to reflect on your REU experiences, focusing on things you learned or gained that you'd consider important to your journey as a scientist.''

\textbf{Exposure to academia}: Several students said their REUs exposed them to academia beyond what they had seen at their home institution. Lexi described coming from a small institution with no graduate programs or research labs and being ``super anti-grad school, super anti-research'' before her REU. She then described her REU as ``kind of the reason [she's] in grad school now.'' Josue, who went to a `state school', said:
\vspace{-1mm}
\begin{quote}\textit{
    There are people that go into academia and STEM in general that know exactly what to look for, like, who to talk to, what to focus on, what skills are important, because they have family members or mentors or things like that, that are in the field. So they kind of have the insider knowledge, I guess. And the REU, for me personally kind of helped bridge that gap, or at least kind of demystify some parts of the grad and research life.
    }
\end{quote}

\textbf{Advisor preferences}: Both students who had and had not done research before their REUs reported that REUs helped them understand what they should look for in advisors and research groups, especially when applying to graduate school. Taylor wanted to attend an REU outside of her college because, while she had done research at her home university, she 11didn't love the culture [there]... both the physics department and... the university community.'' But at her REU, she she said she ``liked the research community'' and later said she wanted to go to grad school in a place ``where I feel like I belong.'' Jordan, who had done very little physics research at the time, said their REU mentor was ``harsh'' and eventually explained that they felt ``nervous for three months straight'',  but that they had ``a lot of fun... with the [other] people I was working with.'' They said that their REU gave them ``a sense of... what do I want to look for in a research group? And how do I avoid \textit{that?}'' (referring to their mentor).

\textbf{Skills and qualifications}: For almost all the students, REUs provided tangible benefits to their scientific career, in skills and/or support for their applications to graduate school. Students felt that doing research at all strengthened them as graduate candidates but also gave them general skills and attitudes related to research. Students like Alejo learned field-specific techniques and knowledge that they used in their future career, though some students noted that none of the field-specific skills they learned were useful to them later on, because they planned to engage in a different field than the one at their REU.
Sebastian was taught to code for the first time, while Taylor received important exposure to the field in which she now works, and both considered their experiences pivotal to their career. However, both noted that they were not able to publish a paper, which Sebastian believed ``makes you less competitive as a grad school applicant''.



\vspace{-0.4\baselineskip}
\subsection{\label{sec:Poly}Barriers to REU attendance}

We asked students, ``[For each REU] talk me through the reason for your decision, and any challenges and barriers you encountered during that process.''

\textbf{Financial status}: Many students commented on financial challenges they encountered in attending their REUs, but what is most striking is that all of them were able to overcome those challenges. Multiple students indicated that their program gave them their first payment later than when they arrived at the REU site. Josue was especially affected by this: 
\begin{quote}
    \textit{We had to live in [REU location] for a week without, like the first half of our stipend. And so I was there and I was broke. And I asked the lady... ``So like, when do we get our paychecks?'' and she was like, ``Sometime on Thursday or Friday,'' and I'm like, ``It's Saturday. Where do I eat?'' And she's like, ``You could, like, go to the grocery store and get like, you know, sandwiches, sandwich bread, and have PB\&J.''}
\end{quote}
Two students noted that they had to pay rent for their lease at home while they attended the REU--for Jordan, that ate deeply into their REU paycheck, while Taylor's parents covered her summer lease so she could attend her REU. Another student said he was paid less at his REU than he would have been at his university, but that he took the pay cut in order to attend. 

While these students had the financial privilege necessary to manage these costs, they are also the students who made it to graduate school. It is reasonable to suggest that many who we do not see in graduate school might have had an insufficient financial safety net to overcome barriers like these.

\textbf{Separation from loved ones}: The potential distance between an REU site and a student's usual home was a factor in some students' decisions because it affected their ability to see their loved ones. One student who ``didn't realize how selective [REUs] are'' said that he and his romantic partner went to different colleges and wanted to be together in their home city during summer--so he only applied to the REU in that city. Another student selected an REU in a neighboring state over an REU several states away because they could take a train to their home state to be with their partner. Yet another student said that they missed their sister's high school graduation in order to attend their REU and called this a ``challenge,'' but did not appear to consider it a barrier to attendance.

\textbf{Response to barriers}: The most striking revelation was the way many students discussed their experience when they were accepted to only one REU. Jordan said, 
\vspace{-1mm}
\begin{quote}
    \textit{I think for me, I was like, just incredibly relieved when I got my acceptance. And so it was kind of a no brainer in terms of like, well, this is what I'm going to go and do this summer and, like, whatever challenges that brings, I'll work through them. Because this is something I really want on my resume, so I can get into grad school.}
\end{quote}
This quote was the most intense expression of this sentiment, but it appeared in other ways. Taylor, who stated that she had the option to do research at her home university, said that ``I think if I hadn't gotten into the [first choice program], I would have ended up going to [second choice program], because it was still better than nothing.'' Another student said that, when COVID hit and multiple REUs that he got accepted to were cancelled, he was ``scampering to find something'' and attended a virtual REU ``for the obvious reason that I wanted to have that on my resume.'' Evidently, students felt strongly that REUs were `important'--to their resume, to their career, and to getting into graduate school. Whether or not REUs actually are important, students who got into only one felt strongly that they needed to get there.

Getting into only one REU was not an ultimate deciding factor for everyone. Nora was willing to give up her study abroad trip to Italy in order to attend her REU, but as a trans woman, she was afraid that she would be ``treated terribly for the whole time'' and needed to confirm that the REU would be accepting and inclusive before she made her decision. If they were not, even though she ``didn't really have any other good options lined up'', she would have stayed home, ``picked up some random internship'', and dealt with ``transphobia at home'' rather than go somewhere without a safety net. She noted, however, that her ability to find other options was a result of being at a major research institution, and it is possible that a student in a less privileged position than Nora would feel no choice but to attend an emotionally unsafe REU that they got accepted to. Another student who got into one REU had also applied for ``on the surface more appealing'' internships, none of which he got into. But he said that ``since I didn't really know if I wanted to go into grad school or not, I wasn't like, bummed to accept the REU.'' Evidently, this student also saw REUs as relevant or even important to the potential career path of graduate school.

REUs have been consistently and widely communicated to students as an important opportunity for their careers as physicists. The students we spoke with believed the same and were intent on attending their REUs. But, given that these were students who made it to graduate school, absent from our sample could be anyone who faced challenges significant enough that they could not attend an REU despite desiring to. We suspect that such a circumstance could be common, but further research is required to investigate the possibility.

\textbf{In-person preference}: Alaee, et al. suggested in Ref. \cite{alaee2022} that students facing geographical constraints would be benefited by remote REUs. In principle, this would mitigate many barriers that underprivileged students may face, but it also comes with a cost: multiple interviewees said they took remote REUs as a last resort or said that they rejected a remote REU in favour of an in-person REU. Marie said that she rejected an REU in her field because it was online and she ``really wanted to work in a lab in person,'' suggesting that students value the laboratory experience. Additionally, interviewees highly valued REUs with social engagement opportunities, and Taylor indicated that she favored her in-person REU over her online REU. Future research should consider whether the most important benefits of REUs, especially for minoritized students, require an in-person experience.
\vspace{-0.6\baselineskip}
\subsection{\label{sec:mentorship}Relationships between REU mentors and students}

We asked students, ``Briefly, how would you characterize your relationship with your REU mentor?'' In what follows, students discuss their faculty advisors.

\textbf{Mentors' expectations}: Students reported widely varying expectations that their mentors had for them. Moreno's mentor initially gave him ``a very small project,'' which due to his background he finished in ``two weeks,'' and afterwards his mentor gave him ``an actual project.'' In contrast, while Josue was a student early in his physics career, his mentor gave him the choice between three research projects he didn't understand, and when Josue picked one at random, he reported that his mentor said ``I was hoping you'd pick that one. That was the hard one.'' When Josue proved unable to complete a task his mentor considered simple, the mentor stopped Josue while he was explaining his difficulties, told Josue to complete the task, and dismissed him with no further training. Other students reported that their mentors were very hands-on in training them: Sebastian said his mentor ``prioritized... development as a professional'' rather than ``new science,'' and in the process Sebastian learned coding skills that enabled his future career, but he also ``never got a paper out.''

There is a dilemma posed by the short duration of REUs and the perceived need for students to obtain deliverables for their resume. Mentors may have a major hand in the resolution of this dilemma for a student. Future research could study how mentors perceive their REU students, not only as researchers contributing to their mentor's project but also as future professionals obtaining much-needed training.

\textbf{How students see mentors}: A striking result from the focus groups relates to how students interacted with mentors when they were relatively new to research mentorship. Multiple interviewees were unsure how to approach their mentors, whether to ask their mentor questions, and what their mentors wanted them to do. Abby said that she had received too little direction from her mentor, but said ``But also, I could have said that to her.'' Nora's mentor rarely reached out to her, but she reported the following interaction when he finally did:
\vspace{-2mm}
\begin{quote}
    \textit{He's like, ``Why didn't you reach out more for help, if you had all these questions?'' And I was like, ``Well, the times that I did try to reach out, you weren't really available, but maybe I probably should have persevered harder.''}
\end{quote}
Elena said,
\vspace{-2mm}
\begin{quote}
    \textit{I didn't really interact with [my mentor] that much. ... I was also like, pretty quiet. ... I always found it kind of hard to ask questions, because I didn't want to ask stupid questions, but maybe I could have asked more questions from her. ... I always felt very anxious and nervous to talk to my PI.}
\end{quote}
The power dynamic between mentor and mentee may drive experiences like these, especially for minoritized or inexperienced students who do not know what they are `allowed' to say to a research mentor. Future research could study how REU students see their mentors and what tools mentors may use to open up communication with their mentees.

\vspace{-0.4\baselineskip}
\section{\label{sec:discussion}Conclusions \& Future Work}

We interviewed 12 graduate students who attended REUs as undergraduates to understand how REUs can support students, particularly minoritized students and students with few research opportunities. Many participants reported benefits from their REUs that match the NSF's goals: students from small colleges and students with limited research experience got skills they might not have developed at their college, and students with knowledge gaps about academia learned valuable information for the first time. As graduate researchers, these students represent clear successes of REU goals, but as relatively affluent students, they do not represent the demographic we hope to also see in a highly-ranked graduate program like this: students who are socioeconomically underprivileged. Many interviewees reported financial barriers attending REUs that they were able to overcome with their socioeconomic privilege. This could be an even larger issue for the low-income students we did not find in our sample. Finally, students reported widely varying experiences with their mentors, but two common themes arose: (1) mentors can provide crucial training for inexperienced students and (2) students do not know what freedoms they have with their REU mentors, what they can say, what questions they can ask, and so on.

These preliminary results inform our future work, which will generate recommendations for REU directors and the NSF. To learn how REUs can be optimized for students with limited research experience we will interview REU students and site directors to find out the best practices. To study barriers, we will survey and then interview undergraduates nationally, searching especially for students who got accepted to REUs and could not attend or who wanted to apply for REUs but chose not to for any reason. Finally, we will further study the relationship between mentors and students in REUs to characterize what role REU mentors can play in enculturating inexperienced students as confident physics researchers.

\acknowledgments{
We would like to acknowledge REU site directors Sathya Guruswamy and Daniel Enrique Serrano for their invaluable insights into REU programs. This work was partially funded by NSF Grant No. 2012147.
}

\bibliography{gradfocusgroups.bib}

\end{document}